\documentclass[aps,prb,reprint,superscriptaddress]{revtex4-2}

\usepackage{graphicx}
\usepackage{xcolor}
\usepackage{hyperref}
\usepackage{tikz}

\begin{document}

\def\lsco{La$_{1-x}$Sr$_{x}$CoO$_3$}
\def\lco{LaCoO$_3$}
\def\lao{LaAlO$_3$}
\def\etal{{\it et al.}}
\def\tc{\ensuremath{T_c}}
\def\den{\ensuremath{\Delta\varepsilon{_{1{\rm N}}}}}

\def\col{\color{blue}}



\title{
	\boldmath
Optical signatures of strain-induced ferromagnetism in LaCoO$_3$ thin film\unboldmath}


\author{F. Abadizaman}
\affiliation{Department of Condensed Matter Physics, Faculty of Science, Masaryk University, Kotl\'a\v{r}sk\'a 2, 611 37 Brno, Czech Republic}

\author{M. Kiaba}
\affiliation{Department of Condensed Matter Physics, Faculty of Science, Masaryk University, Kotl\'a\v{r}sk\'a 2, 611 37 Brno, Czech Republic}

\author{A. Dubroka}
\email{dubroka@physics.muni.cz}
\affiliation{Department of Condensed Matter Physics, Faculty of Science, Masaryk University, Kotl\'a\v{r}sk\'a 2, 611 37 Brno, Czech Republic}
\affiliation{Central European Institute of Technology, Brno University of Technology, 612 00 Brno, Czech Republic}

%
%
%


\date{\today}

\begin{abstract}
Using spectroscopic ellipsometry, we studied the optical conductivity of \lco\ with various degrees of strain. The optical response of the compressively strained \lco\ film is qualitatively similar to the one of the unstrained \lco\ polycrystalline sample and exhibits redistribution of the spectral weight between about 0.2 and  6~eV, which is most likely related to the thermal excitation of the high-spin states. The optical response of the ferromagnetic tensile strained film exhibits clear signatures due to the ferromagnetic state. 
Below the Curie temperature $\tc=82$~K,  the spectral weight is 
transferred with the increasing temperature from low energies between 0.2 and 3.3~eV to energies between 3.3 and 5.6~eV. The temperature dependence of the low-energy spectral weight between 0.2 and 3.3~eV can be understood in the framework of the high-spin biexciton model of Sotnikov and Kune\v{s} as corresponding to the variation of the concentration of high-spin states that are stabilized below \tc. The magnitude of redistribution of spectral weight due to the formation of the ferromagnetic state is sizable. We estimate that it corresponds to a lowering of the kinetic energy of 13~meV per Co ion, which is about two times $k_B\tc$. The latter shows that the saving of the kinetic energy is important and may be the leading energy contribution in the formation of the ferromagnetic phase.

\end{abstract}
\pacs{xxx}
\keywords{keywords }
\maketitle

\section{Introduction}

LaCoO$_3$  has been studied extensively for its unusual magnetic and electronic properties. At low temperatures below 50~K, bulk LaCoO$_3$ exhibits a nonmagnetic insulating behavior with a small optical gap of about 0.2~eV~\cite{Tokura1998,Jeong-scientificreport}.  In an intermediate temperature range between about 50 -- 400~K, it exhibits paramagnetic properties while preserving the insulator properties with a reduced resistance compared to the low-temperature values~\cite{Tokura1998}.  At temperatures above 500~K, the insulator-to-metal 
transition occurs~\cite{Tokura1998}. 
The magnetic properties of \lco\ can be altered by strain. It was observed that the tensile strain induces ferromagnetic (FM) order while  \lco\  remains insulating~\cite{Fuchs2007}. The microscopic mechanism of this FM state is likely qualitatively different from the double-exchange mechanism of the FM state occurring in conducting doped cobaltites~\cite{FuchsPRL2013,Fris2018}, and its nature is a topic of an ongoing debate~\cite{Fujioka2013,Biskup2014,Mehta2015,Feng2019,Sotnikov2020}. 

The unusual electronic and magnetic properties of bulk \lco\ are caused by the specific electronic structure where several spin states are nearly degenerate. It is generally accepted that the unusual behavior at the intermediate temperature range is caused by thermal excitation of higher spin states compared to the low-temperature low-spin (LS) ground state, with the Co electronic configuration $t_{2g}^6$. 
There has been a long debate over which spin state is excited at higher temperatures, whether these are dominantly intermediate-spin (IS) states, $t_{2g}^5e_g^1$, or high-spin (HS) states, $t_{2g}^4e_g^2$~\cite{Degroot1990,Korotin1996,Zobel2002,Ishikawa2004,Yan2004,Haverkort2006,Ropka2003,Podlesnyak2006,Merz2010,Krapek2012}. 
This debate was recently advanced by a joint theoretical and experimental work reporting a pronounced dispersion of IS excitations~\cite{Sotnikov2016,Wang2018,Hariki2020}. This led to a model of thermally excited HS excitons viewed as a biexciton consisting of tightly bound (on the same Co atom) IS excitons with different orbital character. 
In this approach, both IS and HS states are essential in the understanding of the low-energy dynamics of \lco. The energy of HS biexciton is about 20~meV at low temperatures, and its energy significantly increases with temperature~\cite{Hariki2020}.

The strain-induced FM state in \lco\ is usually induced by a substrate with a small lattice mismatch in epitaxial films. The Curie temperature (\tc)  of \lco\ films deposited on (100) oriented LSAT or SrTiO$_3$ substrates is about  80-85~K~\cite{Fuchs2008, Herklotz2009, Mehta2009,Rata2010,Choi2012,Mehta2015} and the \tc\ can reach up to 94~K  in films deposited on (110) oriented LSAT substrates~\cite{Fujioka2013}.
Several aspects potentially important for the mechanism of the FM state were reported. Lattice distortion with propagation vector (1/4,-1/4,1/4) was observed~\cite{Fujioka2013},  an important role of the oxygen vacancies was suggested~\cite{Biskup2014,Mehta2015} and microscopic inhomogeneities of the FM  state were observed~\cite{Feng2019}. Recently, in the framework of HS biexciton model~\cite{Hariki2020}, it was proposed by Sotnikov and Kune\v{s}  that the strain-induced ferromagnetism originates from the interaction between HS states via virtual IS states~\cite{Sotnikov2020}.

In this paper, we would like to contribute to the discussion about the mechanism of the strained induced FM state by examining its optical response. We study the optical properties of unstrained \lco\ (polycrystalline sample) and tensile and compressively strained \lco\ films using ellipsometry in the energy range of the  interband transitions between 0.2 and 6.5~eV. 
Ellipsometry is an established technique that allows the determination of the optical response with a high sensitivity and reproducibility. The temperature-dependent optical response in a wide energy range allows to map the essential redistribution of electronic spectral weight due to the electronic/magnetic transitions, which reflects the changes in the underlying electronic structure~\cite{Basov2011}. 
We observed that the compressive strain does not qualitatively change the optical response. In contrast, in the FM tensile strained film, the optical response is significantly altered and exhibits clear signatures due to the formation of the FM state. We determine the energy range of the FM-related redistribution of the spectral weight and quantify its magnitude.

\section{Experiment}
\label{Experiment}

In this report,  \lco\ samples with different degrees of strain were studied: polycrystalline \lco\ (no strain), \lco\ thin film deposited on LaAlO$_3$ substrate (compressive strain) and \lco\ thin film deposited on 
 (LaAlO$_3$)$_{0.3}$(Sr$_2$TaAlO$_6$)$_{0.7}$ (LSAT) substrate (tensile strain). The thin films were grown using pulsed laser deposition and 
 annealed $in\ situ$ at the deposition temperature of 650~$^\circ$C under 10 Torr oxygen pressure to decrease oxygen vacancy concentration~\cite{Alineason}.
The lateral dimension of thin films is 10$\times$10 mm$^2$, and the thickness of the films was determined using X-ray reflectivity (Rigaku Smartlab) to be about 22~nm. Ellipsometry measurements were performed on the samples and the bare substrates with Woollam VASE ellipsometer (0.6 - 6.5~eV) and Woollam IR-VASE ellipsometer (0.1 - 0.6~eV) in the temperature range from 300 to 7~K. The measurements were repeated several times and were found reproducible. The optical constants of the thin film were obtained at each measured energy from the ellipsometric angles $\Psi$ and $\Delta$ using the standard model of coherent interferences in a thin film on a substrate~\cite{HandbookElli} without the involvement of the Kramers-Kronig relations. The measurements of the magnetic moment were performed using a vibrating sample magnetometer (Quantum Design Versalab).

\section{Data analysis and discussion}

\begin{figure*}
	\centering
	\vspace*{-0.1cm}
	\hspace*{-0.5cm}
	\includegraphics[width=14cm]{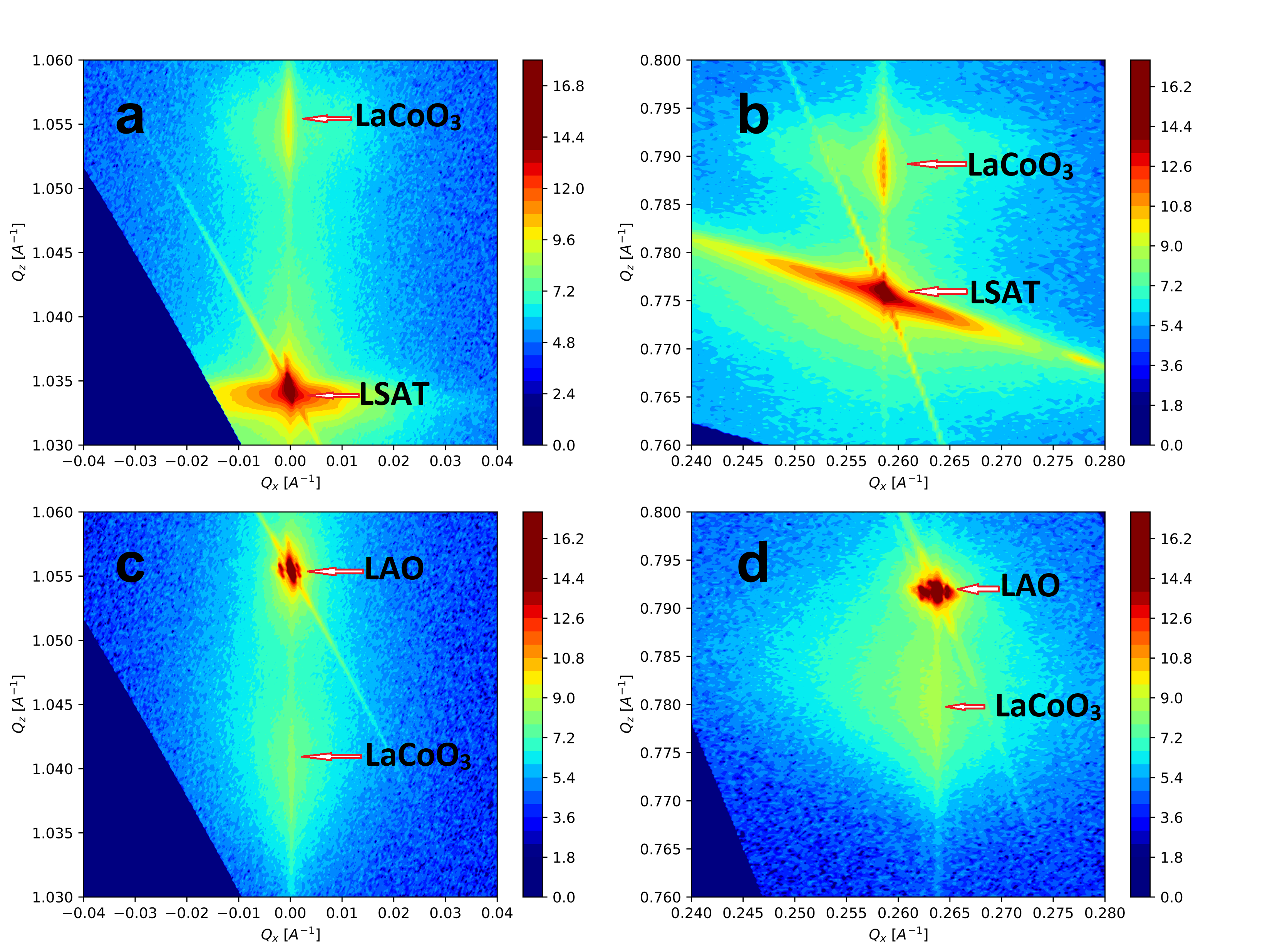}
	\vspace*{-0.2cm} 
 	\caption{X-ray reciprocal space maps of \lco\ film grown on LSAT substrate near the symmetrical (004) diffraction (a) and near the asymmetrical (103) diffraction (b). The diffractions due to the substrate and \lco\ are denoted by the arrows.  Analogical maps are shown for the  \lco\ film grown on \lao\ (LAO)  substrate in panels (c) and (d).}
	\label{RSM}
\end{figure*}

We have probed the structural properties of our films using X-ray diffraction. 
The reciprocal space maps 
of the \lco\ film deposited on  LSAT substrate (\lco/LSAT) measured near the symmetrical (004) diffraction and near the asymmetrical (103) diffraction are shown in Fig.~\ref{RSM}(a) and (b), respectively. Analogical maps for the  \lco\ film deposited on  \lao\  substrate (\lco/\lao) are shown in  Fig.~\ref{RSM}(c) and (d).
They exhibit strong maxima due to the substrate diffraction and distinct maxima due to the \lco\ diffraction at the same $Q_z$ values, depicting that the films are epitaxial and fully strained. The lattice parameters of the samples were determined from the positions of the \lco\ diffractions, and the results are shown in Table~\ref{Lattice}, including the in-plane and out-of-plane values of film strain obtained from the lattice constants. 
The \lco/LSAT film has a tensile strain of 1.07~\%, whereas the \lco/\lao\ film exhibits a compressive strain of -0.99~\%.

\begin{figure}
	\centering
	\vspace*{-0.1cm}
	\hspace*{-0.5cm}
	 \includegraphics[width=8cm]{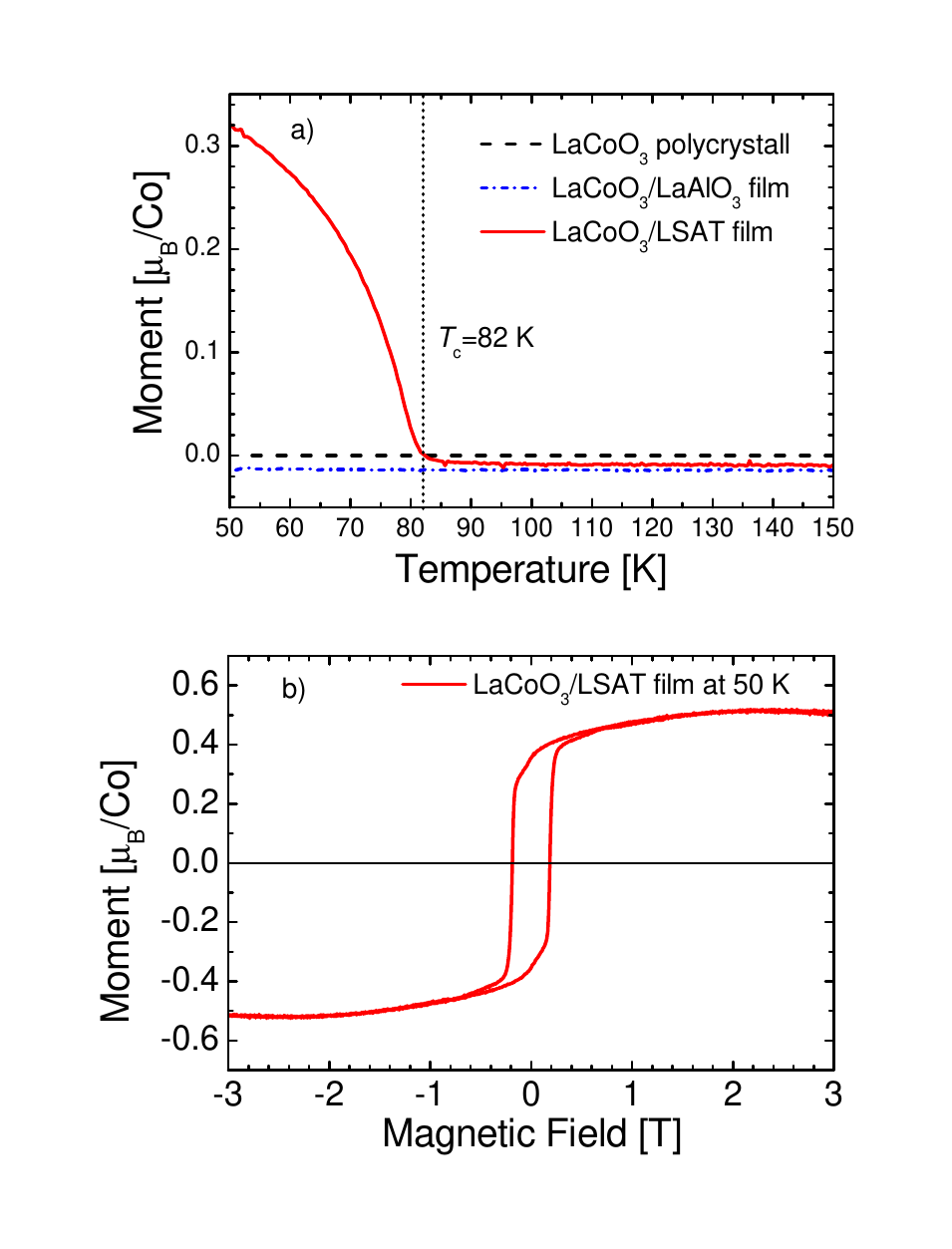}
	\vspace*{-0.2cm}
	\caption{a) Magnetic moment as a function of temperature of \lco\ polycrystalline sample and the \lco/\lao\ and \lco/LSAT films, measured in a magnetic field of 10~mT.
 b) Hysteresis loop of the \lco/LSAT film acquired at 50~K.
 }
	\label{Moment}
\end{figure}

Figure~\ref{Moment}(a) shows the temperature dependence of magnetization of all samples measured in a magnetic field of 10~mT parallel to their surface. The \lco/LSAT film exhibits an onset of FM phase with \tc\ of about  82~K, whereas the other samples are non-magnetic in the measured temperature range. The value of the saturated magnetic moment of the \lco/LSAT film at 50~K is obtained from the hysteresis loop shown in Fig.~\ref{Moment}(b) and amounts to $\mu_s$ = 0.5~$\mu_B$/Co which is in a good agreement with Refs.~\cite{Rata2010} and \cite{Choi2012}. \\\par

\begin{table}[b]
\caption{
Lattice parameters of samples determined by X-ray diffraction. For the bulk samples (polycrystalline \lco\ and the substrates), the lattice parameter corresponds to the bulk pseudocubic lattice constant whereas for the films, it denotes the out-of-plane lattice constant.
}
\begin{tabular}{ l c c c}
 Sample & lattice constant (\AA) & $\varepsilon_{\perp}$ (\%) & $\varepsilon_{\parallel}$  (\%) \\  
 \hline\hline
polycrystalline \lco & 3.825 &  &  \\  
\lao\ substrate & 3.787 &  &  \\
 LSAT substrate & 3.866 &  &  \\
\lco/\lao\ film   & 3.853 & 0.73 & -0.99\\
\lco/LSAT  film & 3.792 & -0.86 & 1.07\\ 
 \hline\hline
 \label{Lattice}
\end{tabular}
\end{table}

\begin{figure*}[t]
	\centering
	\vspace*{-2cm}
	\hspace*{-2cm}
	\includegraphics[width=22cm]{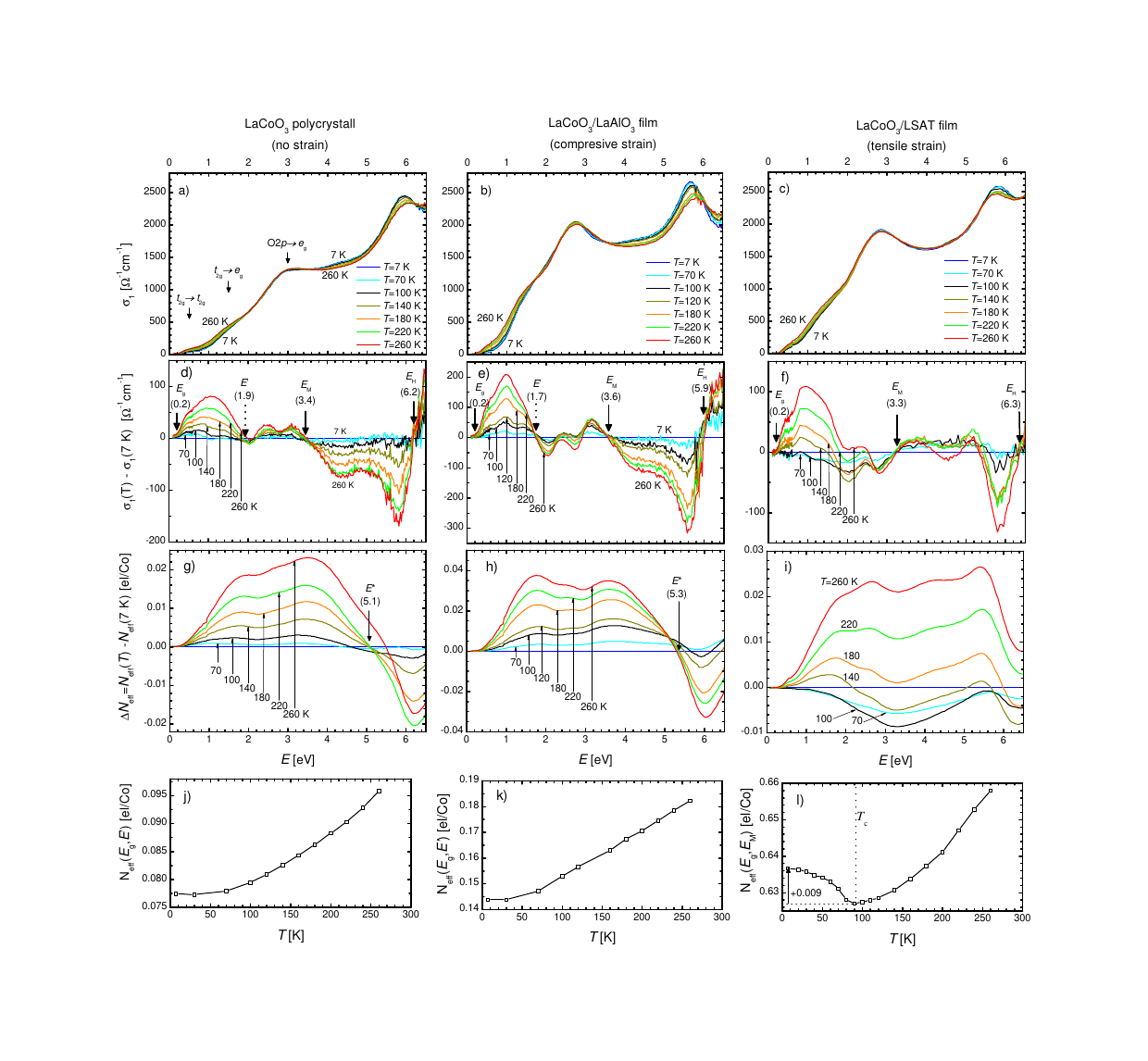} 
	\vspace*{-2.5cm}
 \caption{An overview of the optical data. The left, middle, and right columns correspond to the unstrained \lco\ polycrystall, compressively strained \lco/\lao\ thin film, and tensile strained \lco/LSAT film, respectively.
The top row (a)-(c):  the real part of the optical conductivity $\sigma_1(E,T)$ as a function of the energy $E$ and temperature $T$. 
The second row (d)-(f): $\sigma_1(T)-\sigma_1(7~{\rm K})$. 
The third row (g)-(i): $\Delta N_{\rm eff}(E)=N_{\rm eff}(0.1~{\rm eV},E,T)-N_{\rm eff}(0.1~{\rm eV},E,T=7~{\rm K})$, see Eq.(\ref{Neff}), as a function of the high energy cutoff $E$.
Panels (j) and (k) show the temperature dependence of $N_{\rm eff}(E_g,E')$ and panel (l) shows $N_{\rm eff}(E_g,E_M)$.
}
	\label{S1}
\end{figure*}

An overview of the optical data is presented in Fig.~\ref{S1}. 
The left, middle, and right column corresponds to the \lco\ polycrystall,  the \lco/\lao\ film, and the \lco/LSAT film, respectively.  
The top row presents the real part of the optical conductivity $\sigma_1(E)$ as a function of energy $E$ of the incident photons. Overall, the spectra of all samples have similar shapes.  
They exhibit an optical gap between 0.2 to 0.3~eV and several interband transitions at higher energies that were interpreted as due to  Co $t_{2g}\rightarrow t_{2g}$  transitions (centered at 0.5 eV), Co $t_{2g}\rightarrow e_{g}$ transitions (centered at 1.5 eV) and O 2\textit{p} 
$\rightarrow$ Co $e_{g}$ transitions (centered at 3~eV)~\cite{Jeong-scientificreport}, as marked in Fig.~\ref{S1}(a) by arrows. There is additionally a band near 5.8~eV that corresponds to transitions involving La orbitals since its intensity decreases with increasing Sr doping in  \lsco~\cite{Fris2018}.    
  
The optical data of  strongly correlated electronic materials are often discussed in terms of the effective concentration of charge carriers (or spectral weight)
that is a measure of  absorption of  the electromagnetic radiation in an energy interval ($\hbar\omega_L$, $\hbar\omega_H$):
\begin{equation}
N_{\rm eff}(\omega_L,\omega_H) = \frac{2mV}{\pi e^2} \int_{\omega_L}^{\omega_H} \sigma_1(\omega)d\omega, 
\label{Neff}
\end{equation}
where $m$ is the electron mass, $V$ is the volume of the unit cell, and $e$ is the elementary  charge~\cite{Basov2011}. 
The amount of the spectral weight and the energy intervals where it is redistributed are often evaluated because they reflect the underlying electronic structure and correlations. 
We recall that the optical sum rule states that the total spectral weight  $N_{\rm eff}(0,\infty)$ is equal to the total concentration of charge in a sample, and, consequently, it is temperature independent apart from the effects of thermal expansion that are usually very small and can be neglected.

To visualize the redistribution of spectral weight with temperature, the second row of Fig.~\ref{S1} shows the real part of the optical conductivities relative to the one measured at the lowest temperature of 7~K, $\sigma_1(T)-\sigma_1(7~{\rm K})$. 
Figure~\ref{S1}(d) presents the relative conductivity of   \lco\  polycrystall and depicts that there are several so-called isosbestic energies (energies where the conductivity is essentially temperature-independent) that we denote $E'$, $E_{M}$ and $E_{H}$. 
The largest changes of the spectral weight with increasing temperature involve a decrease between the energies $E_{M}=3.4$~eV and $E_{H}=6.2$~eV and an increase at lower energies between $E_g=0.2$~eV and $E'=1.9$~eV. 
Additionally, there is a smaller temperature dependence of conductivity between $E'$ and $E_{M}$. In the view of the sum rule, we can interpret these findings as a transfer of spectral weight with increasing temperature from high energies between $E_{M}$ and $E_{H}$ to lower energies between $E_{g}$ and $E_{M}$. 
Similar redistribution of optical spectral weight in bulk \lco\  was previously observed by Tokura~\etal\cite{Tokura1998} where it was reported that with increasing temperature, the low-energy spectral weight below 1.4~eV increases on the expense of spectral weight at higher energies. Note that a similar redistribution of the spectral weight induced by a laser pulse was observed with femtosecond ellipsometry~\cite{Zahradnik2022}. The reported crossing point was found to be about 2.1~eV, close to the  $E'= 1.9$~eV found in this report. In the context of 
the HS biexciton model~\cite{Hariki2020}, the observed redistribution of spectral weight is most likely related to the thermal excitation of the high-spin biexcitons and corresponding changes in the occupation of $t_{2g}$ and $e_g$ orbitals.

The relative conductivity of the compressively strained \lco/\lao\ film is shown in Fig.~\ref{S1}(e). Qualitatively, the spectra are similar to those of the polycrystalline \lco\ shown in Fig.~\ref{S1}(d) with slightly different values of the characteristic energies $E'=1.7$, $E_{M}=3.6$ and $E_{H}=5.9$~eV and overall larger magnitude of the changes. Particularly, the spectra exhibit the same redistribution of spectral weight with increasing temperature from high energies between $E_{M}$ and $E_{H}$ to lower energies between $E_{g}$ and $E'$. 

The third row of Fig.~\ref{S1} displays the relative effective concentration of charge carriers with respect to 7~K, $\Delta N_{\rm eff}(E)=N_{\rm eff}(0.1~{\rm eV},E,T)-N_{\rm eff}(0.1~{\rm eV},E,T=7~{\rm K})$, as a function of the high-energy cut-off 
$E=\hbar\omega_H$. The low energy cut-off was chosen to be $\hbar\omega_L=0.1$~eV, right below the optical gap. There is thus a small missing spectral weight due to phonons, but it is negligible compared to the spectral weight of the interband transitions.  
The intersection of $\Delta N_{\rm eff}(E)$ with zero denotes energy below which the redistributions of spectral weight are compensated. 
For the polycrystalline  \lco\ and for the temperature of 220~K, such a zero crossing occurs near $E^*=5.1$~eV, see Fig.~\ref{S1}(g),  and for the compressively strained \lco, it occurs near $E^*=5.3$~eV, see Fig.~\ref{S1}(h). These values of $E^*$ indicate that the spectral weight of the pronounced dip near 5.8~eV in Figs.~\ref{S1}(d) and (e) is not transferred with decreasing temperature to lower energies but to energies above $E_H$. 

The qualitative similarity between the unstrained polycrystalline \lco\ and  the compressively strained \lco/\lao\ film can also be seen in the temperature dependence of the low-energy spectral weight $N_{\rm eff}(E_g,E')$
shown for \lco\ polycrystal 
in Fig.~\ref{S1}(j) and for \lco/\lao\ film in Fig.~\ref{S1}(k). 
Both exhibit a monotonically increasing trend with temperature, which likely corresponds to the thermal population of HS states. 
The similarity of the redistribution of the spectral weight between unstrained polycrystalline \lco\ and the compressively strained \lco/\lao\ film indicates that the compressive strain does not induce a significant modification of the electronic structure of \lco. The only significant difference is quantitative --- the changes with temperature in the compressively strained \lco\  are about twice larger than those of the unstrained \lco. This likely corresponds to a lower activation energy of the HS states in compressively strained \lco\  compared to the unstrained \lco.

Figure~\ref{S1}(f) displays the relative conductivity of the FM \lco/LSAT film. Apparently, the temperature dependence shows qualitative differences with respect to the previous cases shown in Fig.~\ref{S1}(d) and (e).
For example, the isosbectic point near the middle of the measured range at $E_{M}=3.3$~eV has the opposite signature, i.e., the conductivity with increasing temperature decreases below $E_{M}$ in contrast to Fig.~\ref{S1}(d). 
The origin of the differences becomes clear from the temperature dependence of the low energy spectral weight $N_{\rm eff}(E_g,E_M)$ shown in Fig.~\ref{S1}(l), which exhibits a clear onset at \tc\  with the square root behavior typical for the temperature dependence of an order parameter of a second order phase transition. The optical response exhibits clear signatures due to the formation of the FM phase. 

\begin{figure}[t]
	\centering
	\vspace*{-1cm}
	\hspace*{-0.5cm}
	\includegraphics[width=9cm]{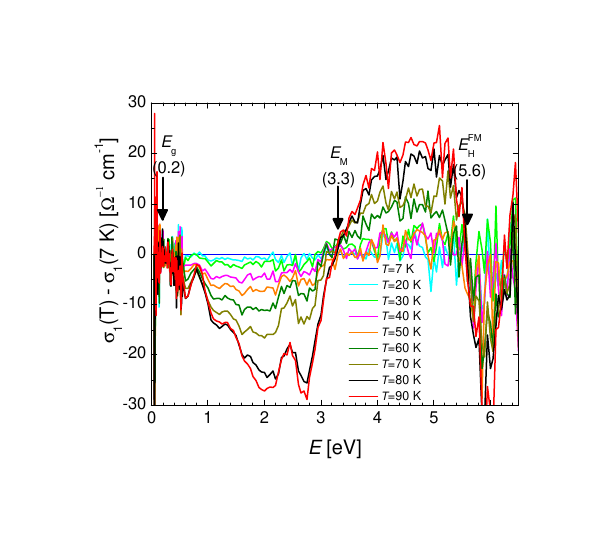}
	\vspace*{-2cm}
	\caption{ The real part of the optical conductivity  of 
 the ferromagnetic \lco/LSAT film with respect to 7~K, $\sigma_1(E,T)$-$\sigma_1(E,7~{\rm K})$, below and near $\tc=82$~K.}
	\label{RELS1FM}
\end{figure}

Figure~\ref{RELS1FM}(a) displays in detail temperature dependence of the  relative optical conductivity $\sigma_1(T)-\sigma_1(7~{\rm K})$ of the \lco/LSAT film only below and near \tc. Since in this temperature range, the temperature changes due to non-magnetic processes are weak, the majority of conductivity changes correspond to the formation of the FM phase. The spectra exhibit
isosbestic points $E_{M}=3.3$~eV and $E^{\rm FM}_{H}= 5.6$~eV as denoted by the arrows. The temperature dependence forms a butterfly-like shape with the isosbestic point $E_{M}$ in the center with two 
``wings" with roughly a similar area between $E_{g}$--$E_{M}$
and $E_{M}$--$E^{\rm FM}_{H}$. 
The spectral weight is transferred with increasing temperature from the low-energy wing into the high-energy wing; 
therefore, the spectral weight redistribution has the opposite trend to the one observed in the unstrained \lco\ shown in Fig.~\ref{S1}(d). 
Note, however, that the redistribution of FM-related spectral weight involves significantly different energy scales than those observed in the non-magnetic samples. In the polycrystalline \lco\ and the compressively strained \lco/\lao\ film, the majority of the spectral weight redistribution at low energies occurs between 0.2 and 1.7-1.9~eV, see Figs.~\ref{S1}(d) and (e). In contrast, in the FM \lco/LSAT film, the low-energy wing spans between 0.2-3.3~eV with the maximum in the interval between 2 and 2.8~eV where the conductivity changes in the non-magnetic samples are only minor. 
We think this sizable difference reflects the FM interaction of the HS states that is absent in the non-magnetic samples. We believe that this sizable change in the involved energy scale is potentially important and should be explained by a theory aiming at a full understanding of the strained-induced FM state in \lco.

Figure~\ref{S1}(l) displays the temperature dependence of the effective concentration of charge carriers $N_{\rm eff}(E_g,E_M)$ in the interval of the low-energy ``wing".  The  amount of the spectral weight redistributed due to the formation of the FM state, $N^{\rm FM}_{\rm eff}$, is estimated (as shown by the arrow) as the difference of $N_{\rm eff}(E_g,E_M)$ between the value at 7~K and right above the \tc\ and amounts to 
$N^{\rm FM}_{\rm eff}=+0.009$ elementary charge per Co ion. This estimate neglects the temperature dependence of $N_{\rm eff}(E_g,E_M)$ due to processes unrelated to FM ordering. We believe that they are low, and if taken into account, they would slightly increase the obtained value of
$N^{\rm FM}_{\rm eff}$.
Providing the low-energy spectral weight $N_{\rm eff}(E_g,E_M)$ is proportional to the population of HS states, its temperature variation shown in 
Fig.~\ref{S1}(l) can be qualitatively understood in the framework of the biexciton model of the strained-induced FM state~\cite{Sotnikov2020}. Above \tc, the concentration of HS states is expected to decrease with decreasing temperature as the thermally populated HS states relax into LS states. Below \tc,  HS states get stabilized by the FM interaction, and thus, their concentration starts to increase with decreasing temperature.

The value of $N^{\rm FM}_{\rm eff}$ can be estimated as well from  $\Delta N_{\rm eff}(E)$ shown in Fig.~\ref{S1}(i) where for $T=100$~K, the value of  
$\Delta N_{\rm eff}(E)$ near 3.3~eV yields the value of -0.009. 
Additionally,  $\Delta N_{\rm eff}(T=100~{\rm K})$
amounts to $-0.001$ at  $E^{\rm FM}_{H}=5.6$~eV  and thus almost reaches zero depicting that within the accuracy of about 10\%, the spectral weight of the high-energy wing compensates for the low-energy wing, and thus the spectral weight below \tc\ is essentially conserved below  $E^{\rm FM}_{H}=5.6$~eV. 
In this context, it is interesting to note that at  260~K, $\Delta N_{\rm eff}(E)$  does not exhibit any zero crossing in the measured range, and at 6.5~eV amounts to a sizable value of 0.008. This demonstrates that in the tensile strained \lco, the spectral weight redistribution at temperatures above \tc\ involves transitions at energies above those occurring in the paramagnetic phase of the unstrained polycrystalline \lco, see Fig.~\ref{S1}(g), and reaches energies beyond our measurement limit of 6.5~eV.

Note that the direction (or sign) of the FM-induced spectral weight redistribution is the same as in the double exchange FM state in doped cobaltites~\cite{Fris2018}, where the FM order is driven by the reduction in the kinetic energy caused by the delocalization of the conducting electrons. Obviously, in \lco, the DC conductivity is absent; however, this does not exclude that the ferromagnetic state is driven by a decrease of the kinetic energy on a finite scale.  This motivated us to quantify the reduction of the kinetic energy $\Delta K$ corresponding to the FM-related redistribution of the spectral weight, $N^{\rm FM}_{\rm eff}$. 
As a rough estimate, we use the same formula as in Ref.~\cite{Fris2018}, $\Delta K =- \frac{3\hbar^2}{a_0^2m_0}N^{\rm FM}_{\rm eff}$, where $\Delta K$ is the change of kinetic energy, $\hbar$ is the reduced Plank's constant, and $a_0$ is the lattice constant. 
This relation was derived for the intraband kinetic energy \cite{Basov2005}. Nevertheless, we use it here merely as a phenomenological estimate of the kinetic energy reduction, even in the present case of interband transitions. Using the value $N^{\rm FM}_{\rm eff}=+0.009$ we obtain $\Delta K = -13$~meV. 
Providing the latter estimate is correct, the change of kinetic energy is about two times the thermal energy at \tc, $|\Delta K/{k_BT_c}| \sim 2$, which shows that the reduction of the kinetic energy is an important quantity that may play a leading role in the mechanism of ferromagnetism in strained \lco. Note that since the penetration depth in \lco\ in the whole measured range is larger than the thickness of the films (about 22~nm), our optical measurements represent an average response of the films. Our findings, therefore, show that the strain-induced FM state is pronounced and probably occurs in the whole or large portion of the film. 
Note additionally that the value of $\Delta K/{k_BT_c} \sim 2$ is essentially the same as the value found for the double-exchange ferromagnetism 
in doped cobaltites~\cite{Fris2018}, again depicting the importance of the magnitude of the FM-related spectral weight redistribution observed in \lco/LSAT thin film.

\section{Summary}
Using spectroscopic ellipsometry, we have measured the optical conductivity of \lco\ with various degrees of strain. The optical response of the compressively strained \lco\ film grown on \lao\ substrate is qualitatively similar to that of the unstrained \lco\ polycrystalline sample. They both exhibit, with increasing temperature, a transfer of spectral weight from high energies between about 3.5 and 6~eV to lower energies, mostly between 0.2 and 1.9~eV. This redistribution of spectral weight is most likely related to the thermal excitation of the high-spin states. 

The optical response of the ferromagnetic tensile strained \lco/LSAT film exhibits clear signatures due to the ferromagnetic state. 
Below the Curie temperature $\tc=82$~K,  the spectral weight is 
transferred with the increasing temperature from low energies between 0.2 and 3.3~eV to high energies between 3.3 and 5.6~eV. 
The sizable spectral range difference of the spectral weight redistribution at low energies with respect to the relaxed \lco\ (3.3~eV instead of 1.9~eV)  most likely reflects the ferromagnetic interaction of the high-spin biexcitons and may be of interest for theoretical predictions. The temperature dependence of the low-energy spectral weight between 0.2 and 3.3~eV can be understood in the framework of the high-spin biexciton model as being due to the variation of the concentration of high-spin states that with lowering temperature are stabilized below \tc. The amount of spectral weight redistributed due to the formation of the ferromagnetic state is sizable and corresponds to 0.009 elementary charge per Co ion. We estimate that it is equivalent to the saved kinetic energy of $\Delta K =$~13~meV. The ratio of $\Delta K$ with respect to ${k_BT_c}$ is about 2, which shows that the reduction of kinetic energy is an important quantity that may be a leading factor in the formation of the ferromagnetic state.

\begin{acknowledgements}
We thank  O. Caha, J. Chaloupka, A. Hariki, J. Kune\v{s}, D. Munzar,  and A. Sotnikov for the fruitful discussions. We acknowledge the usage of data measured by  P. Fri\v{s}. We acknowledge  the financial support by the project
Quantum materials for applications in sustainable technologies,
CZ.02.01.01/00/22\_008/0004572, the MEYS of the Czech Republic under the
project CEITEC 2020 (LQ1601), by the Czech Science Foundation (GACR) under Project No. GA20-10377S and CzechNanoLab project LM2023051 funded by MEYS CR for the financial support of the measurements/sample fabrication at CEITEC Nano Research Infrastructure.
\end{acknowledgements}

\bibliographystyle{apsrev4-2}
\bibliography{bibliographyCobaltites}

\end{document}